\documentclass[prl,twocolumn,showpacs,floatfix,amsmath,amssymb,superscriptaddress]{revtex4}
\usepackage{latexsym}
\usepackage{graphicx}
\usepackage{times}
\usepackage{color}
\usepackage{amsmath}
\usepackage{dcolumn}
\usepackage{latexsym,amsmath,amssymb,bm,euscript}
\begin{document}

\title{Giant topological insulator gap in graphene with 5d adatoms}

\author{Jun Hu}
\affiliation{Department of Physics and Astronomy, University of California, Irvine, California 92697}
\author{Jason Alicea}
\email[]{aliceaj@uci.edu}
\affiliation{Department of Physics and Astronomy, University of California, Irvine, California 92697}
\author{Ruqian Wu}
\email[]{wur@uci.edu}
\affiliation{Department of Physics and Astronomy, University of California, Irvine, California 92697}
\author{Marcel Franz}
\affiliation{Department of Physics and Astronomy, University of British Columbia, Vancouver, BC, Canada V6T 1Z1}


\begin{abstract}
{Two-dimensional topological insulators (2D TIs) have been proposed as platforms for many intriguing applications, ranging from spintronics to topological quantum information processing.  Realizing this potential will likely be facilitated by the discovery of new, easily manufactured materials in this class.  With this goal in mind we introduce a new framework for engineering a 2D TI by hybridizing graphene with impurity bands arising from heavy adatoms possessing partially filled $d$-shells, in particular osmium and iridium.  First principles calculations predict that the gaps generated by this means exceed 0.2 eV over a broad range of adatom coverage; moreover, tuning of the Fermi level is not required to enter the TI state.  The mechanism at work is expected to be rather general and may open the door to designing new TI phases in many materials.  
} 
\end{abstract}

\maketitle

Topological insulators comprise a class of strongly spin-orbit-coupled, non-magnetic materials that are electrically inert in the bulk yet possess protected metallic states at their boundary \cite{KaneMele,KaneReview,QiReview}.  These systems are promising sources for a host of exotic phenomena---including Majorana fermions \cite{FuKaneMajoranas,MajoranaQSH,BeenakkerReview,AliceaReview}, charge fractionalization \cite{TEC}, and novel magneto-electric effects \cite{Magnetoelectric1,Magnetoelectric2,SpinJosephson1,SpinJosephson2}---and may also find use for quantum computing \cite{KaneReview} and spintronics devices \cite{Nagaosa}.  In some respects two-dimensional (2D) topological insulators are ideally suited for such applications; for example, bulk carriers that often plague their three-dimensional counterparts can be vacated simply by gating.  Experimental progress on 2D topological insulators has steadily advanced recently due largely to pioneering work on HgTe \cite{HgTeQSHtheory,HgTeQSHexpt,HgTeQSHexpt2,HgTeQSHexpt3} (see also Ref.\ \cite{Knez}).  Nevertheless, to realize their full potential systems more amenable to experimental investigations are highly desirable.  In this regard the ability to design new 2D topological insulators from conventional, widely available materials would constitute a major step forward, and many proposals of this spirit now exist \cite{Raghu, Weeks, Tami,Lindner,PRX,OxideInterface,PolarizationTI,MolecularGrapheneTI}.  

Following this strategy, here we introduce a new mechanism for engineering a topological insulator state in graphene---arguably now the most broadly accessible 2D electron system.  Historically, graphene was the first material predicted to realize a topological insulator in seminal work by Kane and Mele \cite{KaneMele}, though unfortunately the gap is unobservably small due to carbon's exceedingly weak spin-orbit coupling \cite{GrapheneSO1, GrapheneSO2, GrapheneSO3, GrapheneSO4, GrapheneSO5}.  Reference \cite{PRX} revived graphene as a viable topological insulator candidate by predicting that dilute concentrations of heavy In or Tl adatoms dramatically enhance the gap to detectable values of order 0.01 eV.  Essentially, these adatoms mediate enhanced spin-orbit interactions of the type present in the Kane-Mele model \cite{KaneMele} for pure graphene.  

Our approach here also relies on hybridizing graphene with dilute heavy adatoms, though the underlying physics is entirely different and can not be understood in terms of an effective graphene-only model.  Rather, we will show using density functional theory that certain adatoms---specifically Os, Ir, Cu-Os dimers, and Cu-Ir dimers---form spin-orbit-split impurity bands that hybridize with graphene's Dirac states in such a way that a highly robust topological insulator regime emerges.  In fact here it is more appropriate to view the adatoms as the dominant low-energy degrees of freedom, with \emph{their} coupling effectively mediated by graphene; from this perspective this mechanism represents the inverse of that invoked in Ref.\ \cite{PRX}.  

Numerous practical advantages arise in this scheme.  The topological insulator gaps are extremely large---typically exceeding 0.2 eV---and take on nearly the full \emph{atomic} spin-orbit splitting for the adatoms.  Such values reflect more than an order-of-magnitude enhancement compared to the gaps induced by In or Tl, and are competitive with the largest gap predicted for any topological insulator.  Somewhat counterintuitively, these gaps are remarkably insensitive to the adatom concentration, taking on comparable values at least over coverages ranging from $\sim2\% - 6$\%.  In the case of Os adatoms and Cu-Ir dimers, the Fermi level also naturally resides within the topological insulator band gap.  This eliminates perhaps the most serious challenge with In and Tl, both of which substantially electron-dope graphene even at quite low coverages.  These features suggest that the observation of a topological insulator state in graphene may be within reach.  

\begin{figure}
\includegraphics[width = 8.5 cm]{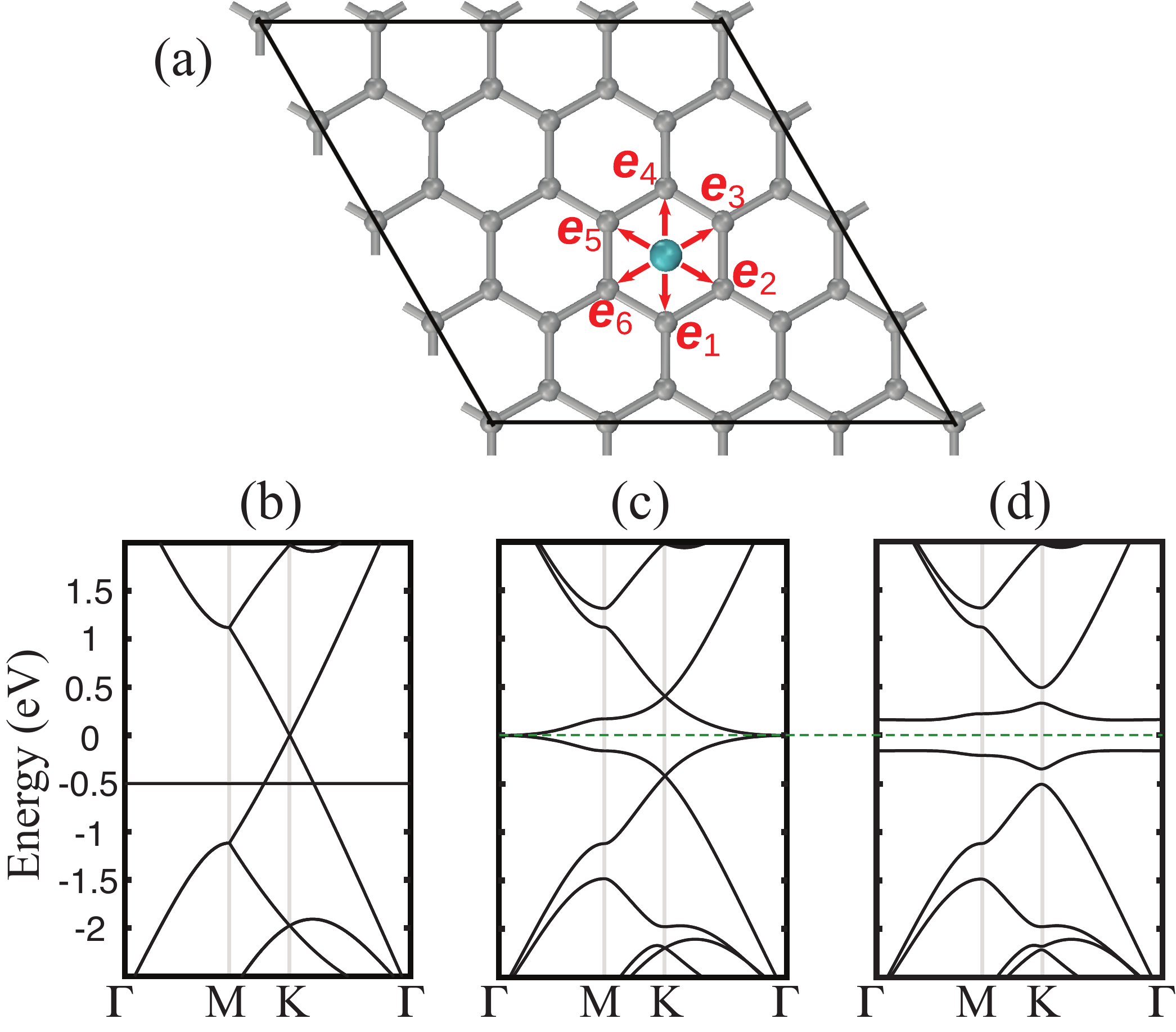}
\caption{(a) $4\times 4$ supercell employed to simulate periodic H-site adatoms (cyan) at 6.25\% coverage.  (b)-(d) Corresponding tight-binding band structures calculated with graphene hopping strength $t = 2.7$ eV and adatom on-site energy $\epsilon = -0.5$ eV. The adatom-graphene hopping $t_c$ and adatom spin-orbit coupling $\Lambda_{so}$ are given by (b) $t_c = \Lambda_{so} = 0$, (c) $t_c = 1.5$ eV, $\Lambda_{so} = 0$, and (d) $t_c = 1.5$ eV, $\Lambda_{so} = 0.2$ eV.  When the Fermi level sits at the dashed green line, $\Lambda_{so} \neq 0$ generates a topological insulator gap given approximately by the \emph{atomic} adatom spin-orbit splitting.  
}
\label{BandStructureFig}
\end{figure}

We first elucidate the mechanism uncovered here using a tight-binding model that exposes the physics in a very transparent manner.  Consider $5d$ adatoms residing at positions ${\bf R}$ located at `hollow' (H) sites in graphene as in Fig.\ \ref{BandStructureFig}(a).  For simplicity we retain only the $d_{xz}$ and $d_{yz}$ adatom states since these comprise the most important orbitals in our first-principles calculations.  (Recall that $d_{xz/yz}$ orbitals arise from $L^z$ orbital angular momentum $m = \pm 1$ states.)  We then model the composite system by a Hamiltonian $H = H_g + H_a + H_c$ \cite{PRX}.  The first term allows nearest-neighbor hopping for graphene:
\begin{equation}
  H_g = -t\sum_{\alpha = \uparrow,\downarrow}\sum_{\langle {\bf r r'}\rangle}(c_{{\bf r}\alpha}^\dagger c_{{\bf r}'\alpha} + H.c.),
  \label{Hg}
\end{equation}
where $c_{{\bf r}\alpha}^\dagger$ adds an electron with spin $\alpha$ to honeycomb site ${\bf r}$.  The second encodes couplings for the adatoms,
\begin{eqnarray}
  H_a &=& \sum_{\bf R}\bigg{[}\sum_{\alpha = \uparrow,\downarrow}\sum_{m = \pm 1} \epsilon f_{m{\bf R}\alpha}^\dagger f_{m{\bf R}\alpha} 
  \nonumber \\
  &+& \sum_{\alpha,\beta = \uparrow,\downarrow}\Lambda_{so}(f_{1{\bf R}\alpha}^\dagger s^z_{\alpha\beta} f_{1{\bf R}\beta} - f_{-1{\bf R}\alpha}^\dagger s^z_{\alpha\beta} f_{-1{\bf R}\beta})\bigg{]}.
  \label{Ha}
\end{eqnarray}
Here $f_{m{\bf R}\alpha}^\dagger$ fills the adatom $d$-orbital at position ${\bf R}$ with magnetic quantum number $m = \pm 1$ and spin $\alpha$, $\epsilon$ sets the orbital energies relative to graphene's Dirac points, $\Lambda_{so}$ represents spin-orbit coupling, and $s^z$ is a Pauli matrix.  Finally, $H_c$ hybridizes the adatoms with graphene.  To express this term it is convenient to define vectors ${\bf e}_j$ that point from an adatom to the six surrounding carbon sites [see Fig.\ \ref{BandStructureFig}(a)].  One can then construct linear combinations $C_{m{\bf R}} = \frac{1}{\sqrt{6}}\sum_{j = 1}^6 e^{-i\frac{\pi}{3}m(j-1)} c_{{\bf R} + {\bf e}_j}$ that carry angular momentum $m$ and write \cite{PRX}
\begin{equation}
  H_c = -t_c\sum_{\bf R}\sum_{\alpha = \uparrow,\downarrow}\sum_{m = \pm1} (i C_{m{\bf R}\alpha}^\dagger f_{m{\bf R}\alpha} + H.c.).
  \label{Hc}
\end{equation}

Let us now specialize to a periodic adatom arrangement characterized by the $4\times 4$ supercell shown in Fig.\ \ref{BandStructureFig}(a), with one adatom per cell (this corresponds to $6.25\%$ coverage).  Figure \ref{BandStructureFig}(b) illustrates the band structure with $t = 2.7$ eV, $\epsilon = -0.5$ eV, and $t_c = \Lambda_{so} = 0$.  In this limit the adatoms produce a four-fold degenerate flat band, reflecting spin and orbital degeneracy.  For the following discussion the precise location of these adatom states is unimportant, provided they intersect the carbon bands within $\sim 1$ eV of the Dirac points.  Incorporating tunneling between the adatoms and graphene causes the flat bands to disperse, as shown in Fig.\ \ref{BandStructureFig}(c) for $t_c = 1.5$ eV.  Suppose now that the Fermi level resides at the dashed line in Fig.\ \ref{BandStructureFig}(c).  Although the spectrum here exhibits a sizable energy gap near the $K$ point, the system remains metallic due to band touchings at the zone center.  The gapless excitations at zero momentum exhibit the following two crucial properties: $(i)$ they arise from \emph{weakly perturbed adatom orbitals} since at the zone center the nearest carbon bands reside well over 1 eV away in energy, and $(ii)$ they are protected by time-reversal, spatial rotation, and SU(2) spin symmetries that coexist when $\Lambda_{so} = 0$.  Breaking the last of these symmetries by turning on spin-orbit coupling thus produces a bulk gap given nearly by the \emph{atomic} spin-orbit splitting for the adatoms, despite their dilute coverage.  This key point is demonstrated in Fig.\ \ref{BandStructureFig}(d) for $\Lambda_{so} = 0.2$ eV, which yields a 0.32 eV gap that constitutes 80\% of a single adatom's spin-orbit splitting.

\begin{figure}
\includegraphics[width = 8.5 cm]{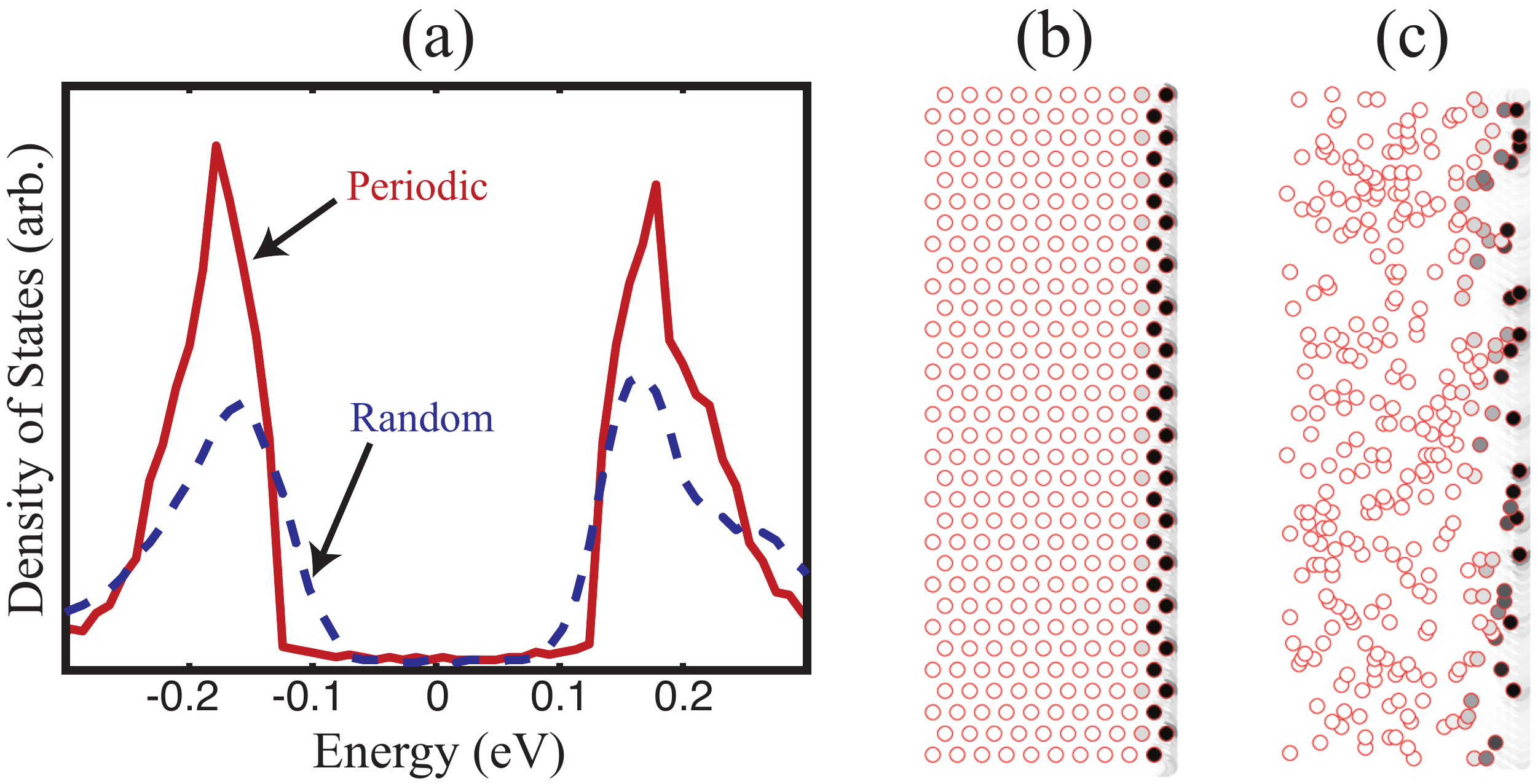}
\caption{(a) Density of states for periodic (solid curve) and random (dashed curve) adatoms at 6.25\% coverage on a graphene strip with armchair edges along $x$ and periodic boundary conditions along $y$.  Parameters are the same as for Fig.\ \ref{BandStructureFig}(c).  The finite density of states within the bulk gap reflects edge states, which remarkably survive even for randomly distributed adatoms.  Examples of edge states for the periodic and random cases respectively appear in (b) and (c).  
}
\label{EdgeStateFig}
\end{figure}

The gap opening indeed drives the system into a topological insulator phase.  Since our Hamiltonian is inversion symmetric this can be readily verified by computing Fu and Kane's formula for the $Z_2$ invariant in Ref.\ \cite{FuKane}.  For additional evidence the solid curve in Fig.\ \ref{EdgeStateFig}(a) plots the density of states (DOS) near zero energy for the same periodic adatom coverage on a graphene strip with armchair edges along $x$ and periodic boundary conditions along $y$.  (Our strip consists of 128 zig-zag `rows' of carbon sites, with 80 sites per row.)  Edge states characteristic of the topological phase produce a finite DOS inside of the bulk gap, and are clearly resolvable in the system size simulated over an energy window of 0.31 eV.  As an example, an edge state with mid-gap energy $E = 0.004$ eV appears in Fig.\ \ref{EdgeStateFig}(b), where circles indicate adatom locations while the shading represents the probability amplitude extracted from the wavefunction.  

Remarkably, the formation of a topological insulator by no means requires the periodic arrangements considered so far.  In fact similar physics arises even for \emph{completely randomly} distributed H-site adatoms.  The dashed curve in Fig.\ \ref{EdgeStateFig}(b) illustrates the DOS for the random case, again at 6.25\% coverage.  Even in our finite system one can easily resolve edge modes within a 0.21 eV energy range that is comparable to the bulk gap for the periodic case; see, \emph{e.g.}, the mid-gap state with energy $E = 0.003$ eV plotted in Fig.\ \ref{EdgeStateFig}(c).  

Next we demonstrate using density functional theory (DFT) that the mechanism described above can be realized in graphene with 5$d$ adatoms, notably Os and Ir.  All DFT calculations were carried out with the Vienna ab-initio simulation package (VASP) \cite{Vasp1, Vasp2} at the level of the local density approximation (LDA) \cite{LDA}, including spin-orbit coupling unless specified otherwise.  Positions of all atoms were fully relaxed using the conjugated gradient method for energy minimization until the calculated force on each atom became smaller than 0.01 eV/{\AA}.  Most results were obtained using the supercell in Fig.\ \ref{BandStructureFig}(a), with one adatom per cell.  For additional details see \footnote{We used the projector augmented wave (PAW) method for the description of the core-valence interactions \cite{Vasp3, Vasp4}.   A vacuum space of 15 {\AA} was adopted to separate the periodic graphene slabs. The two-dimensional Brillouin zone was sampled by a $15\times15$ $k$-grid mesh \cite{Monkhorst}. The energy cutoff of the plane wave expansion was set to 500 eV. }.    Below we report our results for Os and then turn to Ir.  

Since the thermal stability of adsorption structures is relevant for both experiments and applications, candidate adatoms should ideally exhibit large H-site binding energies [defined as $E_b = E({\rm graphene})+E({\rm adatom})-E({\rm adatom/graphene})$] and high diffusion energy barriers [defined as $\Delta E=E_b ({\rm Transition~state}) - E_b ({\rm Ground~state})$].  Osmium satisfies both criteria.  The binding energy for Os at the H site in graphene is 2.42 eV---much larger than the `top' (directly above a C) and `bridge' (above the midpoint of a C-C bond) configurations for which $E_b$ is 1.70 and 1.59 eV, respectively.  Moreover, the calculated diffusion barrier for an Os adatom to diffuse from an H site through the top site is found to equal the difference between the binding energies at these positions, 0.72 eV.  The barrier for diffusion through the bridge site is similarly given by 0.83 eV.  Therefore Os adatoms should be stable over H sites even at room temperature.  By contrast most 3$d$ transition metals have $E_b \sim 1$ eV \cite{CohenAdatoms} and are much more mobile \cite{CohenAdatoms, K.M.Ho}; for example, the diffusion barrier is only 0.40 eV for Co on graphene \cite{PRB.82.045407}.  

\begin{figure}
\includegraphics[width = 8.2 cm]{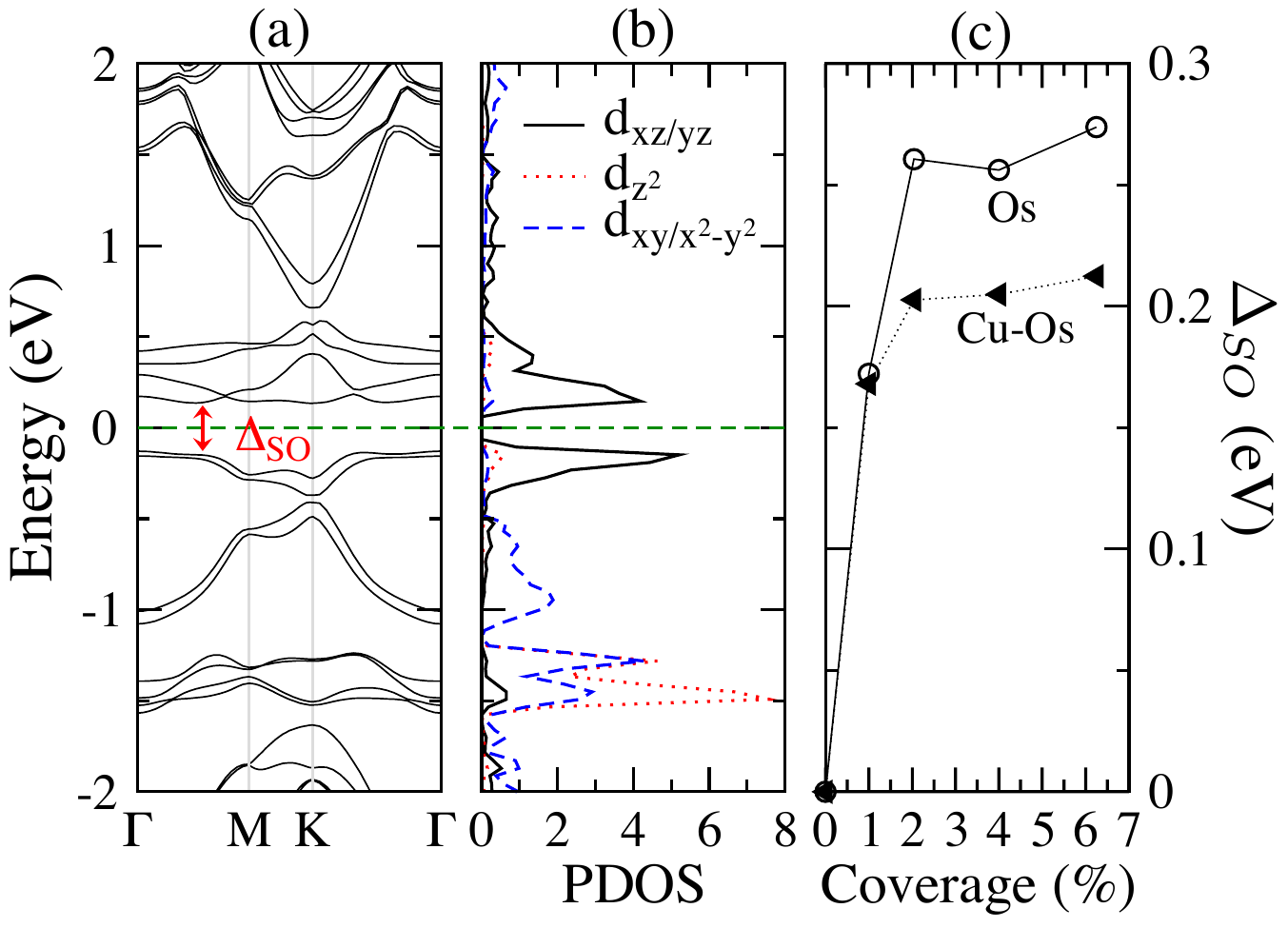}
\caption{(a) First principles band structure for Os on graphene at 6.25\% coverage.  The green dashed line indicates the Fermi level.  (b) Corresponding partial density of states (PDOS) for the Os $5d$ levels.  The large gap $\Delta_{SO}$ visible in (a) arises from hybridization between graphene and the spin-orbit-split $d_{xz/yz}$ orbitals, as in our tight-binding model.  (c) Coverage dependence of the gap for graphene with Os adatoms (circles) and Cu-Os dimers (triangles).  
}
\label{bands}
\end{figure}

Figure \ref{bands}(a) displays the DFT band structure for periodic H-site Os adatoms on graphene at 6.25\% coverage using the $4\times 4$ supercell in Fig.\ \ref{BandStructureFig}(a).  Each Os adatom forms a charge state of $+0.55e$ (based on the Bader charge division scheme \cite{Bader}), indicating that the Os-graphene bonds mix covalent and ionic features.  Clearly these bonds dramatically modify the characteristic Dirac bands at the $K$ point of pure graphene similar to Fig.\ \ref{BandStructureFig}.  Most importantly, Os induces a large band gap $\Delta_{SO}= 0.27$ eV, \emph{right at the Fermi level} given by the green dashed line in Fig.\ \ref{bands}(a).  As in our tight-binding model the gap here results solely from spin-orbit coupling.  (Without spin-orbit interactions a gapless spectrum arises; see the Supplemental Material.)  More precisely, the partial density of states (PDOS) for the Os $5d$ orbitals displayed in Fig.\ \ref{bands}(b) indicates that the gap arises from the hybridization between graphene's $\pi$ states and the spin-orbit-split $d_{xz}$ and $d_{yz}$ adatom orbitals, also as in our tight-binding model.  The PDOS for the $d_{z^2}$, $d_{x^2-y^2}$, and $d_{xy}$ orbitals, by contrast, is concentrated at much lower energies.  Thus the gap-opening mechanism introduced earlier indeed appears in the realistic Os/graphene system.  

The Os-induced gap depends exceptionally weakly on coverage.  To illustrate this important point we performed simulations of graphene with one Os adatom in $5\times 5$, $7\times 7$, and $10\times 10$ supercells (corresponding to coverages of 4\%, 2.04\%, and 1\%).  Circles in Fig.\ \ref{bands}(c) show the DFT-predicted gaps, which remain close to 0.2 eV even at 1\% coverage.   This striking feature is actually rather natural since the local atomic spin-orbit splitting for the Os $d_{xz}$ and $d_{yz}$ orbitals essentially sets $\Delta_{SO}$.

\begin{figure}
\includegraphics[width = 8.5 cm]{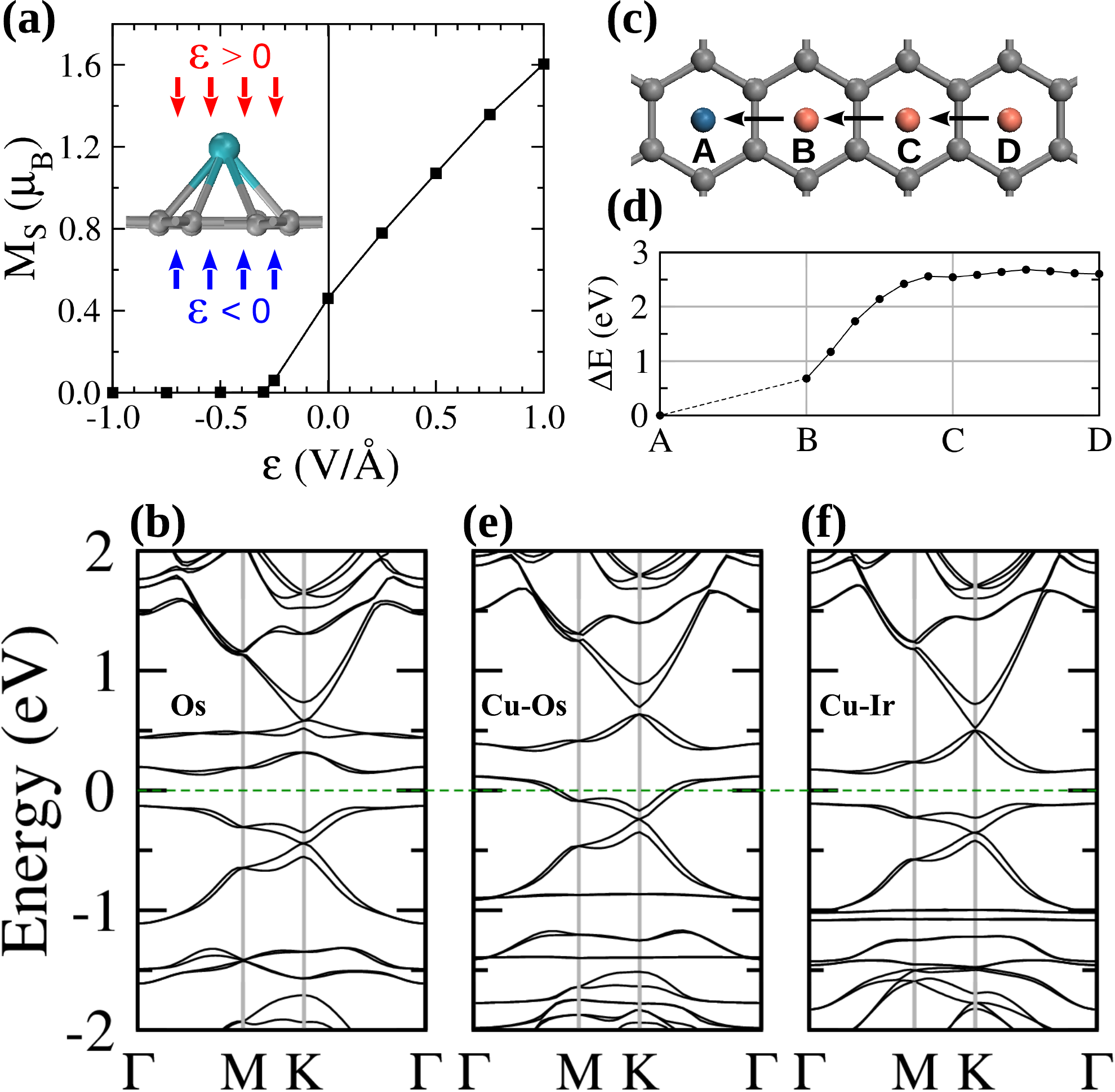}
\caption{(a) Magnetic moment $M_s$ of Os/graphene versus external electric field $\varepsilon$ applied perpendicular to the graphene sheet (see inset for the direction of positive and negative $\varepsilon$).  The moment disappears for $\varepsilon \lesssim -0.3$ V/{\AA}, allowing a true topological insulator phase to appear.  (b) Band structure of Os/graphene with $\varepsilon = -0.5$ V/{\AA}.  Comparing to Fig.\ \ref{bands}(a) one sees that the fields needed to restore time-reversal symmetry modify the band structure very little; in particular a large gap remains at the Fermi level (green dashed line).  (c) Possible diffusion path of a Cu atom beginning from position D and ending above an Os atom at position A.  (d) Energy profile along this diffusion trajectory.  The small diffusion barrier evident in (d) indicates that Cu-Os dimers should form quite efficiently.  (e) Band structure for Cu-Os dimers on graphene at 6.25\% coverage.  Time-reversal symmetry is preserved here even at $\varepsilon = 0$, though the Fermi level now resides in the valence band.  (f) Band structure for Cu-Ir dimers on graphene at 6.25\% coverage.  This system preserves time-reversal symmetry, eliminates the shift in Fermi level, and also supports a large topological insulator gap.   
}
\label{modifying}
\end{figure}

Strictly speaking, a true topological insulator phase does not arise in the DFT simulations described above since Os forms small spin and orbital magnetic moments of 0.45 $\mu_B$ and 0.05 $\mu_B$, respectively.  This produces visible splittings of the bands at the $\Gamma$ and M points corresponding to time-reversal-invariant momenta; see Fig.\ \ref{bands}(a).  One should keep in mind, however, that DFT can sometimes overestimate moment formation.  Nonetheless, even if a moment $M_s$ indeed appears in an experiment, there are practical means by which this can be quenched to zero to reveal a bona fide topological phase \footnote{Actually for some applications such as the pursuit of Majorana fermions, time-reversal symmetry breaking can be a feature rather than a bug; see Refs.\ \cite{MajoranaQSH} and \cite{AliceaReview}}.  One effective approach is to apply an external electric field $\varepsilon$.  Figure \ref{modifying}(a) illustrates that $M_s$ of Os on graphene depends sensitively on $\varepsilon$.  In particular $\varepsilon < 0$ transfers additional charge from Os to graphene and kills the moment for $\varepsilon \lesssim -0.3$ V/{\AA}.  The electric fields required to restore time-reversal symmetry only weakly affect the band structure.  See, for example, Fig.\ \ref{modifying}(b) corresponding to 6.25\% Os coverage with $\varepsilon = -0.5$ V/{\AA}, where a time-reversal-invariant topological insulator appears with a gap $\Delta_{SO} = 0.26$ eV.

Co-doping provides another means to quench the Os magnetic moment. To preserve the main features of the band structure while attracting charge away from Os (as accomplished by a negative $\varepsilon$), co-adsorbates should interact weakly with graphene and exhibit larger electronegativity than Os. Following this guidance, we considered Cu, Ag, and Au in several configurations as described in the Supplemental Material.  Whereas Os repels Ag and Au adatoms, Cu prefers to climb over Os to form a vertical Cu-Os dimer over the H site.  The binding energy $E_b = E({\rm graphene})+E({\rm Cu})+E({\rm Os})-E({\rm Cu-Os/graphene})$ for these dimers is 5.96 eV, higher by 2.50 eV compared to that of well-separated Cu and Os adatoms.  Additionally, Cu more strongly anchors Os to the H site since the binding energy for the vertical dimers over the top (bridge) site is weaker by 1.27 (1.42) eV.  This is not the only relevant criterion, however, since in practice isolated Cu and Os adatoms must be able to dimerize without overcoming substantial energy barriers.  We explored this issue by computing the total energies along the diffusion path depicted in Fig.\ \ref{modifying}(c), where a Cu adatom beginning at position D ends up above an Os at position A \footnote{Here we calculated the Cu diffusion energy barrier without spin-orbit coupling.  The energy is not expected to change significantly, however, with the inclusion of relativistic effects since Cu exhibits relatively weak spin-orbit interactions.  Note also that since Cu is physisorbed on graphene, changes to the trajectory will not appreciably alter the diffusion barrier.}.  Figure \ref{modifying}(d) illustrates the change in energy $\Delta E$ relative to the dimer state for various positions along this trajectory.  The energy barrier for a Cu adatom to diffuse from location D to B is only $\sim 0.08$ eV; once at position B the Cu strongly attracts to the top of Os and forms the vertical dimer \footnote{More precisely, when the Cu is located at position B, DFT predicts that it spontaneously climbs to the top of Os to form the vertical dimer when we allow the atomic positions to relax.  This is why no intermediate points between A and B are shown in Fig.\ \ref{modifying}(d).}.  This suggests that dimer formation ought to proceed quite efficiently.  

Because of the hybridization and charge transfer between the Cu and Os atoms---the Bader charges of Cu and Os are respectively $-0.21e$ and $+0.67e$---DFT predicts that graphene with Cu-Os dimers is nonmagnetic.  The spectrum for (Cu-Os)/graphene at 6.25\% coverage again supports a large topological insulator gap $\Delta_{SO} = 0.21$ eV as evident in the band structure of Fig.\ \ref{modifying}(e).  Moreover, the triangles in Fig.\ \ref{bands}(c) show that this gap exhibits similarly weak coverage dependence as for Os/graphene.  The drawback here, however, is that the Fermi level [green line in Fig.\ \ref{modifying}(e)] now resides in the valence band.  Returning the Fermi level to the insulating regime should be possible with conventional gating techniques, provided one works at low coverage.  

Alternatively, the hole introduced by each Cu-Os dimer can be compensated by replacing Os with Ir, which has one additional electron.  Our calculations show that vertical Cu-Ir dimers also strongly bind to the H-site in graphene without forming a magnetic moment.  Hybridization between Cu-Ir dimers and graphene produces nearly the same band structure as for (Cu-Os)/graphene, but with the Fermi level lying in the band gap.  See the band structure for 6.25\% Cu-Ir coverage in Fig.\ \ref{modifying}(f), where the gap is $\Delta_{SO} = 0.25$ eV.  Additional results for Ir/graphene---which behaves similarly to Os/graphene---can be found in the Supplemental Material.

In summary, we have introduced a mechanism by which graphene covered with heavy adatoms realizes a topological insulator protected by a giant gap comparable to atomic spin-orbit energies, even at exceptionally dilute coverages.  Using DFT we predicted that Os, Ir, Cu-Os dimers, and Cu-Ir dimers all give rise to this mechanism and produce gaps exceeding 0.2 eV at coverages as low as 2\%.  Although our DFT calculations of necessity invoked periodic adatom configurations, our tight-binding simulations indicate that readily observable bulk (mobility) gaps should survive also in the random case relevant for experiments.  These findings are expected to greatly facilitate the realization of a topological insulator phase in graphene-based systems.  We suspect, however, that the mechanism exposed here has much broader applications since (contrary to Ref.\ \cite{PRX}) the physics has nothing to do with the Kane-Mele model specific to graphene.  Hybridizing trivial metals or insulators with heavy-element impurity bands may therefore provide a generic method for designing new topological phases, which would be interesting to investigate in future work.

\acknowledgments{The authors gratefully acknowledge A.\ Damascelli, J.\ Eisenstein, J.\ Folk, E.\ Henriksen, and C.\ Zeng for helpful discussions, as well as C.\ Weeks for performing transport calculations related to this study.  This work was supported by DOE Grant DE-FG02-05ER46237 (JH and RW), the National Science Foundation through Grant DMR-1055522 (JA), the Alfred P. Sloan Foundation (JA), NSERC and CIfAR (MF). }

\section{Supplemental Material}

\subsection{Electronic properties without spin-orbit coupling}

\begin{figure}[t]
\includegraphics[width = 8.5 cm]{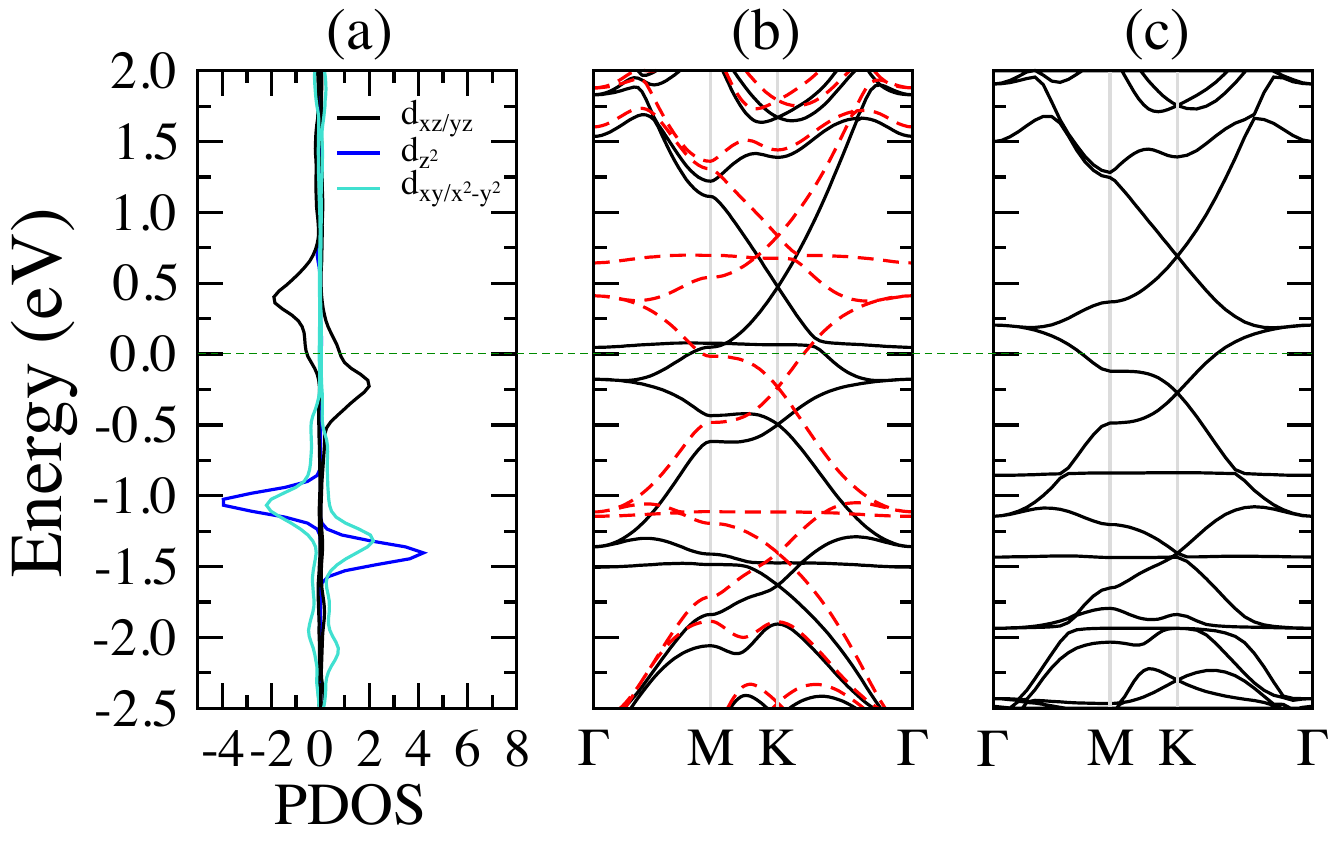}
\caption{Electronic properties of Os/graphene and (Cu-Os)/graphene at 6.25\% coverage \emph{without} spin-orbit coupling. (a) Partial density of states (PDOS) for the Os 5$d$ orbitals in Os/graphene. Positive and negative values on the horizontal axis correspond to the PDOS for the spin majority and minority channels, respectively. The horizontal dashed lines indicate the Fermi level. (b) Band structure for Os/graphene.  The solid and dashed lines respectively indicate majority and minority spin bands, which are widely separated due to strong moment formation in the non-relativistic limit. (d) Band structure of (Cu-Os)/graphene.  In both (b) and (c) the spectrum is always metallic, demonstrating that the gaps found earlier indeed originate from spin-orbit coupling.  
}
\label{figS-band-noSOC}
\end{figure}

To confirm that the band gaps induced by Os and Cu-Os stem from spin-orbit coupling, we performed DFT calculations for these adatoms at 6.25\% coverage in the non-relativistic limit.  For Os/graphene, we find that in the absence of spin-orbit coupling the Os spins polarize much more strongly compared to the relativistic case yielding a magnetic moment $M_s = 1.52 \mu_B$.  The spin splitting is clearly visible in the partial density of states for the Os $5d$ orbitals shown in Fig.\ \ref{figS-band-noSOC}(a), where positive and negative values on the horizontal axis correspond respectively to the majority and minority spin channels.  The band structure for the majority spins (solid curves) and minority spins (dashed curves) appears in Fig.\ \ref{figS-band-noSOC}(b).  Notice that in sharp contrast to the spin-orbit-coupled case, both spin channels are always gapless for any value of the chemical potential (at least over the energy range shown).  Interestingly, (Cu-Os)/graphene remains non-magnetic even in the absence of spin-orbit coupling. From the band structure in Fig.\ \ref{figS-band-noSOC}(c) one clearly sees that here, too, the system remains gapless in the non-relativistic limit.  Thus for both Os/graphene and (Cu-Os)/graphene the gaps indeed originate solely from spin-orbit interactions.  

\subsection{Energies of different Cu-Os co-adsorption configurations}

\begin{figure}[t]
\includegraphics[width = 8.5 cm]{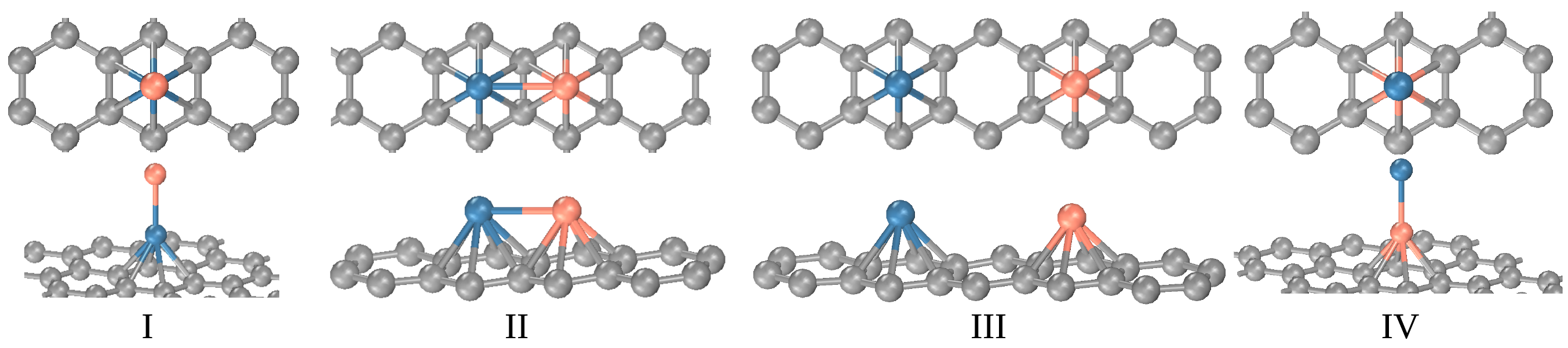}
\caption{ Different configurations for Cu-Os co-adsorption. The dark cyan, coral, and gray spheres represent Os, Cu, and C atoms, respectively.
}
\label{figS-structure}
\end{figure}

In the main text we remarked that Cu and Os co-adsorbates energetically prefer to form vertical dimers over the H-site in graphene.  Here we summarize the evidence supporting this conclusion.  We used DFT, without spin-orbit coupling, to explore the energetics of the various configurations displayed in Fig.\ \ref{figS-structure} (Os, Cu, and C atoms are respectively represented by dark cyan, coral, and gray spheres).  Note that in all cases we fix the Os to the H-site since the binding energy for that position greatly exceeds that for the top or bridge sites.  (Additionally, Cu-Os dimers bind much more weakly to the top and bridge positions.)  The vertical Cu-Os dimer in configuration I exhibits a very large binding energy $E_b = 6.17$ eV.  In contrast the binding energy for configurations III and IV are substantially lower by 2.50 and 2.09 eV, respectively.  Configuration II is unstable and transforms to configuration I after relaxation, indicating a strong tendency toward vertical dimer formation.  As we saw earlier the vertical dimer in (a) is nonmagnetic (with or without spin-orbit coupling), whereas the other configurations are magnetic.

\subsection{Results for Ir/graphene}

\begin{figure}
\includegraphics[width = 8.5 cm]{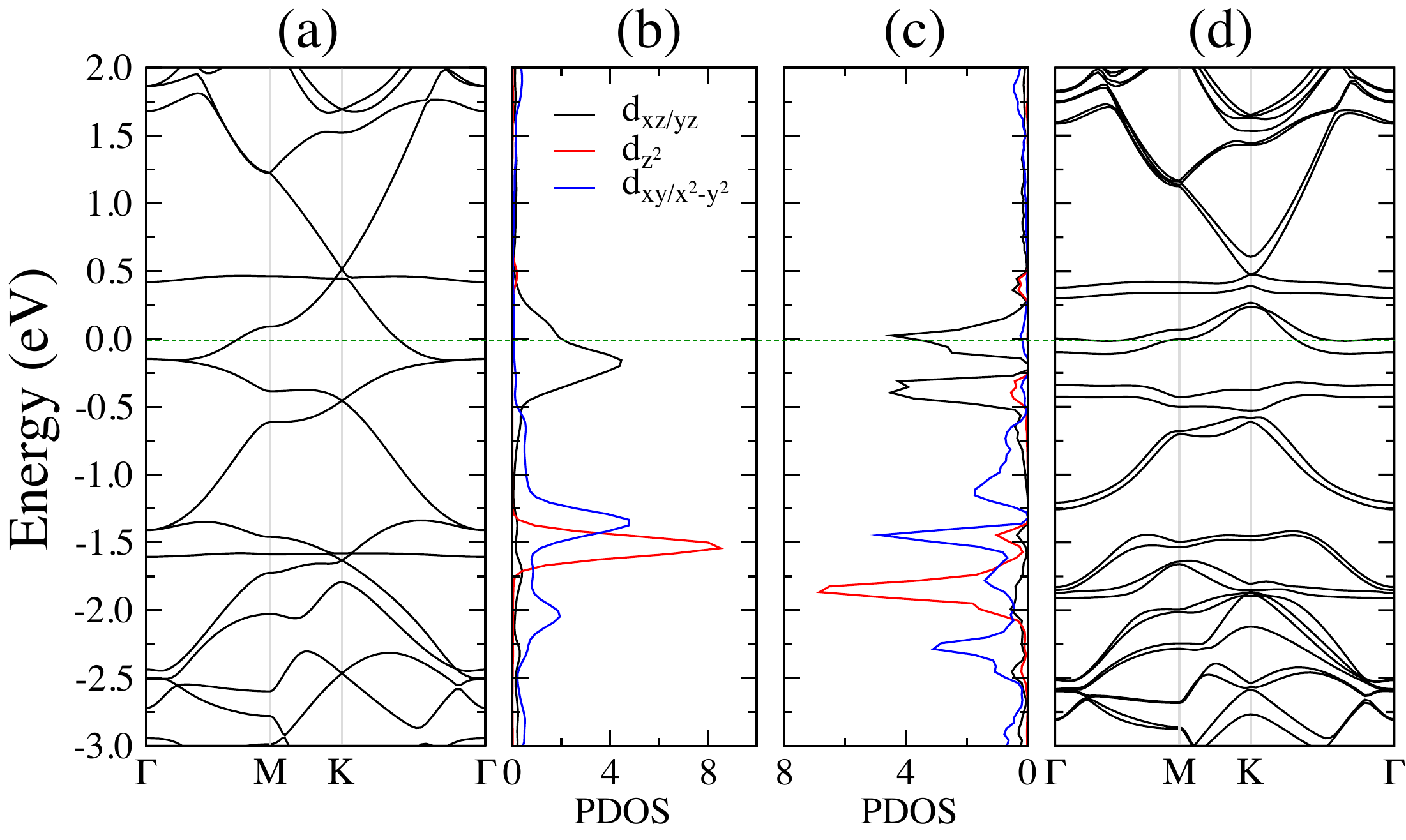}
\caption{(a) Band structure and (b) PDOS for Ir 5d orbitals obtained for Ir/graphene at 6.25\% coverage, \emph{without} SOC.  The green horizontal dashed line denotes the Fermi level. Here the system is nonmagnetic, and the spectrum is again always gapless similar to Figs.\ \ref{figS-band-noSOC}(b) and (c).  In (c) and (d) we display the PDOS and band structure calculated with spin-orbit coupling.  A large spin-orbit-induced gap opens below the Fermi level, though the bands are split slightly due to weak moments formed by Ir in this case.  As discussed in the main text replacing Ir with Cu-Ir dimers returns the Fermi level to within the gap \emph{and} quenches the moment to restore time-reversal symmetry.  
}
\label{figS-Ir-band}
\end{figure}

Finally, we briefly highlight our results for Ir adatoms on graphene.  We found that Ir slightly prefers the H site over the bridge site, the binding energy being 2.17 eV in the former configuration and 2.12 eV in the latter.  The binding energy for Ir at the top site is weakest at 1.94 eV.  All results discussed henceforth thus correspond to H-site Ir adatoms.  Without spin-orbit coupling DFT predicts that Ir/graphene is non-magnetic.  Figures \ref{figS-Ir-band}(a) and (b) illustrate the band structure and partial density of states for the Ir $5d$ levels, calculated at 6.25\% coverage.  Just as for Os and Cu-Os dimers, here too the system remains gapless in the non-relativistic limit.  

\begin{figure}
\includegraphics[width=6cm]{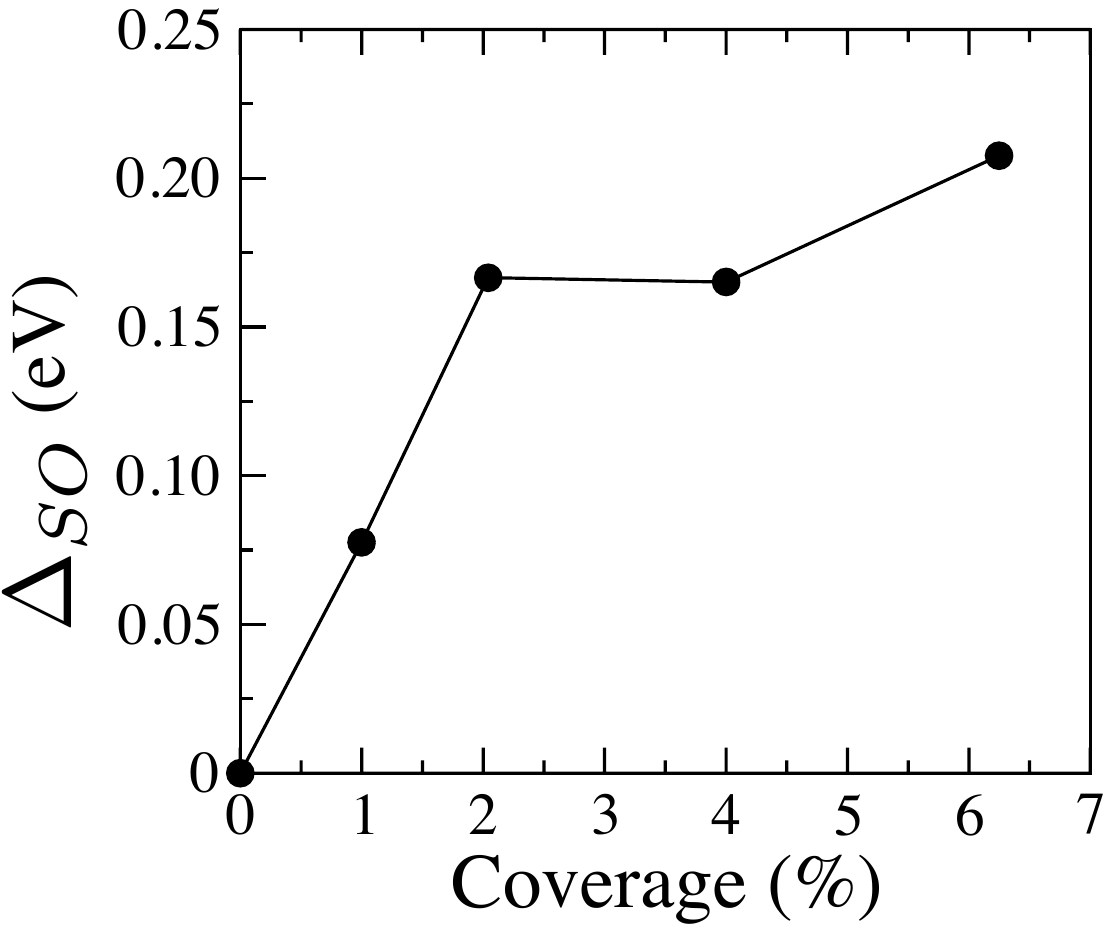}
\caption{Spin-orbit-induced band gap of Ir/graphene as a function of Ir coverage.  Data were obtained using $4\times4$, $5\times5$, $7\times7$ and $10\times10$ graphene supercells that contain one Ir adatom over the H site.
}
\label{figS-Ir-coverage}
\end{figure}

Figure \ref{figS-Ir-band}(d) shows that restoring spin-orbit coupling introduces a huge band gap $\Delta_{SO} = 0.21$ eV below the Fermi level (green dashed line).  Moreover, in Fig.\ \ref{figS-Ir-band}(c) we see that this gap results from hybridization between the Ir $d_{xz/yz}$ orbitals and graphene's $\pi$ bands, similar to Os/graphene and the tight-binding model described in the main text.  Spin-orbit interactions also, however, produce small spin and orbital moments for Ir given respectively by 0.30 $\mu_B$ and 0.13 $\mu_B$; these are responsible for the weak band splittings in Fig.\ \ref{figS-Ir-band}(d).  Coverage dependence of the gap is illustrated in Fig.\ \ref{figS-Ir-coverage}.  As in all other cases studied here, the gap remains quite large down to very dilute Ir concentrations.  Even at 1\% coverage where $\Delta_{SO}$ reduces to 0.08 eV the gap exceeds that induced by In or Tl adatoms \cite{PRX} many times over.  

Observing a true topological insulator phase in Ir/graphene is complicated by the magnetic moments predicted by DFT and the fact that the Fermi level resides in the conduction band.  (One should again keep in mind, however, that DFT may overestimate the tendency for moment formation here.  Furthermore, it may be feasible to gate the system back to the insulating regime by conventional means at low coverages.)  In the main text we described how co-doping with Cu eliminates both potential challenges.  Here we simply wish to note that Cu-Ir dimers carry yet another advantage---they strongly enhance binding to the H-site compared to isolated Ir adatoms, similar to the Os case.  Indeed, the binding energy for Cu-Ir dimers over the H-site is larger by 0.62 eV and 0.79 eV compared to the bridge and top sites, respectively.  By contrast the differences for pure Ir are only 0.05 eV and 0.23 eV as we saw above.

\end{document}